\documentstyle[prd,aps,preprint]{revtex}
\tightenlines
\input epsf.tex
\def\DESepsf(#1 width #2){\epsfxsize=#2 \epsfbox{#1}}
\begin{document}

\draft
%@twocolumnfalse\endcsname
%\preprint{\hbox{CTP-TAMU-48-99}}
\title{Neutralino Proton Cross Sections For Dark Matter In SUGRA and D-BRANE
Models} 

\author{  R. Arnowitt, B. Dutta and Y. Santoso }

\address{
Center For Theoretical Physics, Department of Physics, Texas A$\&$M
University, College Station TX 77843-4242}
\date{May, 2000} 
\maketitle
\begin{abstract}
Neutralino proton cross sections are examined for models with R-parity
invariance with universal soft breaking (mSUGRA) models, nonuniversal SUGRA
models, and D-brane models. The region of parameter space where current
dark matter detectors are sensitive, i.e. $1\times 10^{-6}$ pb, is examined. 
For mSUGRA models, detectors are sampling parts of the parametr space for
tan$\beta\stackrel{>}{\sim}25$. The nonuniversal models can achieve cross
sections that are a factor of 10-100 bigger or smaller then the universal one
and in the former case sample regions tan$\beta\stackrel{>}{\sim}4$. The D-brane
models considered require tan$\beta\stackrel{>}{\sim}15$. The inclusion of CP
violating phases reduces the cross section by a factor of $\sim$ 2-3 (but also
requires considerable fine tuning at the GUT scale). The expected particle
spectra at accelerators are examined and seen to differ for each model. Three
new regions of possible coannihilation are noted.
\end{abstract}

\section{Introduction} Supersymmetric theories with R-parity invariance predict
the existence of a dark matter candidate, the lightest supersymmetric particle
(the LSP). Accelerator and cosmological constraints \cite{falk} then imply that the
LSP is the lightest neutralino, ${\tilde\chi^0_1}$, and that this particle is
mainly gaugino \cite{arnowitt1}. For heavy nuclei, ${\tilde\chi^0_1}$-nucleus scattering
is dominated by the spin independent part of the amplitude where the neutron (n)
and proton (p) amplitudes are approximately equal. This allows one to extract
from the data the spin independent cross section $\sigma_{\tilde\chi^0_1-p}$.
Current dark matter experiments  are now sensitive enough to probe a significant
part of the SUSY parameter space. Thus DAMA and CDMS are sensitive to 
$\sigma_{\tilde\chi^0_1-p}$ in the range of $(1-10)\times 10^{-6}$ pb, and
there will be perhaps a factor of 10 improvement in the near future. We ask here
then what part of
the SUSY parameter space can be tested for the range 
\begin{eqnarray}
0.1\times 10^{-6} {\rm pb}\leq\sigma_{\tilde\chi^0_1-p}\leq 10\times 10^{-6} {\rm pb}.
\end{eqnarray}
We do this by examining the maximum theoretical cross section that lies in this
domain by varying the SUSY parameters (e.g. tan$\beta$, $m_{\tilde\chi^0_1}$,
etc). In the following, we consider three supergravity models: mSUGRA with
universal soft breaking parameters
\cite{kelley,arnowitt2,kane,baer1,arnowitt3,baer2,baer3,barger1,ellis1}, 
SUGRA models with non-universal soft 
breaking  in both the 
Higgs \cite{berezinsky1,berezinsky2,nath1,arnowitt4,bottino3,arnowitt5} and third generation squark/slepton sector
\cite{nath1,arnowitt4,arnowitt5}, and a D-brane model which allows for a specific pattern of 
nonuniversal gaugino, squark and slepton masses \cite{brhlik2}.

In our analysis \cite{aads}, we let the SUSY soft breaking parameters $m_{\tilde g}$
(gluino mass), $m_0$ (scalar mass), $A_0$ (cubic soft breaking mass) and
tan$\beta=<H_2>/<H_1>$ range over the following domain:
\begin{eqnarray}
m_{\tilde g},\, m_0&\leq& 1\,{\rm TeV};\,\, |A_0/m_0|\leq 5;\,\, 2\leq tan\beta\leq 50.
\end{eqnarray}

In order that the analysis be accurate for this domain, we include the
folllowing in the calculations; (i) we run the full 1-loop renormalization group
equations (RGE) from the GUT scale  $M_G=2\times 10^{16}$ GeV to the t -quark
mass $m_t$=175 GeV, iterating to get a consistent SUSY mass spectrum for a fixed
set of GUT scale mass parameters. (ii) L-R mixing in sfermion mass matrices are
included (which is important for large tan$\beta$ for the third squark and
slepton generations). (iii)One loop corrections to the  Higgs mass matrix is
included. (iv) One loop corrections to the mass $m_b$ is included
(which is also important for large tan$\beta$ where it produces a significant
correction to the relation between $\lambda_b$ (the b-Yukawa coupling constatnt)
and $m_b$. (v) QCD RGE corrections are included for contributions dominated by
light quark masses. (vi) leading order (LO) and approximate NLO corrections
\cite{anlauf,baer3} to the $b\rightarrow s\gamma$ decay rate are included. We do not impose
$b-\tau$ unification at $M_G$ (as is done in \cite{bottino3}), as this constraint is sensitive to possible
unknown GUT scale physics.

Various LEP, Tevatron and CLEO accelerator bounds limit the SUSY parameter
space. For the light chargino $\tilde \chi^{\pm}_1$ and light higgs $h$ we require 
\begin{eqnarray}
m_{\tilde \chi^{\pm}_1}&>&94 {\rm GeV};\,m_h>95 {\rm GeV},
\end{eqnarray}
and for the $B\rightarrow X_s\gamma$ branching ratio we use \cite{cleo}:
\begin{eqnarray}
1.8\times 10^{-4}\leq B(B\rightarrow X_s\gamma)\leq 4.5\times 10^{-4}.
\end{eqnarray}
The b-quark mass used is $m_b(m_b)$=4.0-4.5 GeV\cite{PDG} and 
Tevatron bounds \cite{tev} on
$m_{\tilde g}$ and $m_{\tilde q}$ are imposed.

The theoretical analysis determines the ${\tilde\chi^0_1}-q$ scattering cross
section, and in order to relate this to the proton cross section two parameters
$\sigma_0$ and $\sigma_{\pi N}$ enter \cite{ellis2,bottino1}, where
\begin{eqnarray}
\sigma_{\pi N} &=& \frac{1}{2} (m_{u} + m_{d}) \langle p | \bar{u} u + \bar{d} d | p
\rangle\ \\\nonumber
\sigma_{0} &=& \frac{1}{2} (m_{u} + m_{d}) \langle p | \bar{u} u + \bar{d} d -2 \bar{s} s| p
\rangle\ .	\label{eq8}
\end{eqnarray}
 as well as the quark mass ratio $r = {m_{s}}/{{1\over 2}(m_{u}+m_{d})}$.
We use here $\sigma_{\pi N}=65\ \rm{MeV}$, in accord with 
recent analysis\cite{mgo}
and $\sigma_{0}=30\ \rm{MeV}$ \cite{bottino1}. The quark mass ratio is given in \cite{leutwyler}
as r=24.4$\pm$1.5.

Theory allows one to calculate the mean $\tilde{\chi}_{1}^{0}$ relic cold matter
(CDM) of the universe, parametrized by $\Omega_{\tilde{\chi}_{1}^{0}}h^{2}$,
where $\Omega_{\tilde{\chi}_{1}^{0}}=\rho_{\tilde{\chi}_{1}^{0}}/\rho_c$ 
($\rho_{\tilde{\chi}_{1}^{0}}$= mass density of $\tilde{\chi}_{1}^{0}$,
$\rho_c=3 H_0^2/{8\pi G_N}$, $H_0$= Hubble constant) and h=$H_0/(100 km/s
Mpc)$). The combined world average for $H_0$ is \cite{freedman} $H_{0} = (71 \pm 3 \pm
7)\ \rm{km\ s^{-1}\ Mpc^{-1}}\ $. A number of astronomical analyses of matter
density $\Omega_{m} $ gives \cite{dodelson,mohr} $\Omega_{m} \simeq 0.3 $, and using the
baryon density $\Omega_{B} = 0.05$, we assume that
$\Omega_{\tilde{\chi}^{0}_{1}} = 0.25 \pm 0.10\ $. In view of possible
systematic errors that may exist, we chose here a 2 std range for 
$\Omega_{\tilde{\chi}^{0}_{1}}h^{2}$:
\begin{equation}
0.02 \leq \Omega_{\tilde{\chi}^{0}_{1}}h^{2} \leq 0.25\ .	\label{eq4}
\end{equation}
The theoretical formula for $\Omega_{\tilde{\chi}_{1}^{0}}h^{2}$ is given by\cite{jungman}:
\begin{equation}
\Omega_{\tilde{\chi}_{1}^{0}}h^{2} = 2.48 \times 10^{-11} \left(
\frac{T_{\tilde{\chi}_{1}^{0}}}{T_{\gamma}} \right)^{3} \left(
\frac{T_{\gamma}}{2.73} \right)^{3} \frac{N_{f}^{1/2}}{\int_{0}^{x_{f}} dx \langle
\sigma_{ann} v_{rel} \rangle }		\label{eq15}
\end{equation}
where $T_{\tilde{\chi}_{1}^{0}}$ is the freeze out
 temperature $T_{f}$($x_{f} = kT_{f}/m_{\tilde{\chi}_{1}^{0}}$), 
 $N_{f}$ is the number of degrees of freedom at freezeout,
 $(T_{\tilde{\chi}_{1}^{0}}/T_{\gamma})^{3}$ is the reheating factor,
 $\sigma_{ann}$ is the
annihilation cross section, $v_{rel}$ is relative velocity, and 
$\langle \ ...\rangle$ means thermal average.

\begin{figure}[htb]
\bigskip
\bigskip
\bigskip
\centerline{ \DESepsf(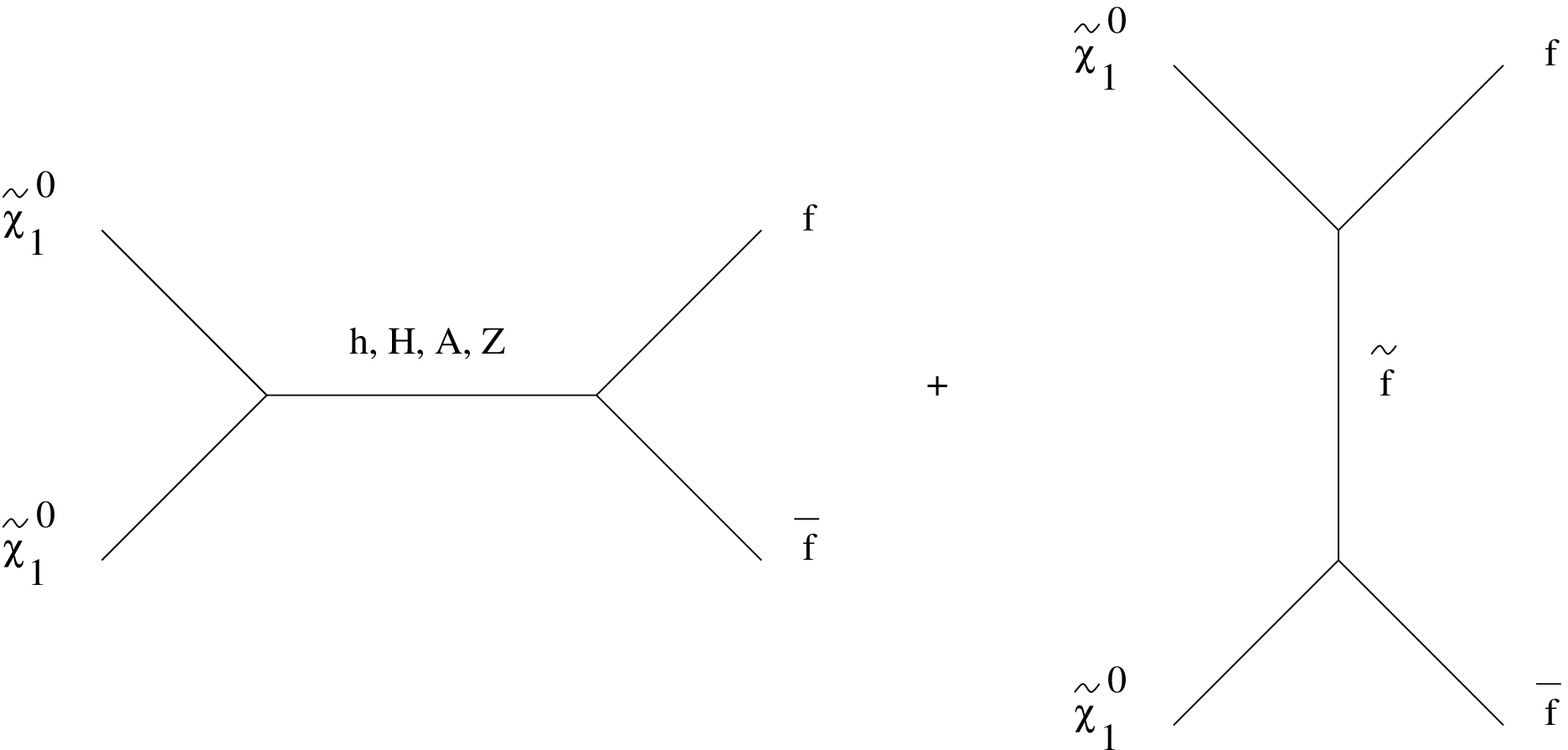 width 13 cm) }
\bigskip
\bigskip
\caption {\label{fig1} Early universe annihilation of
$\tilde{\chi}_{1}^{0}$ through s channel Higgs ($h$, $H$, $A$) and $Z$ poles 
and t channel squark
and slepton ($\tilde{f}$) poles.}
\end{figure}

Eq.(6) puts a strong constraint on the SUSY parameter space and this then affects the
$\tilde\chi^0_1-p$ cross section. From Fig.~\ref{fig1} one sees that (i) $\sigma_{ann}$
decreases with increasing m$_0$ and $m_{\tilde\chi^0_1}$ 
(thus increasing $\Omega_{\tilde\chi^0_1}h^2$) and (ii) if 
$2 m_{\tilde\chi^0_1}$ is near (but below) $m_h$, $m_H$, $m_A$ rapid
annihialtion occurs through s -channel pole (thus decreasing
$\Omega_{\tilde\chi^0_1}h^2$). While LEP has now eliminated most of the
parameter space where $2 m_{\tilde\chi^0_1}\simeq m_h$, both $H$ and $A$ become
light for large tan$\beta$ making this effect latter significant in that domain.

\begin{figure}[htb]
\centerline{ \DESepsf(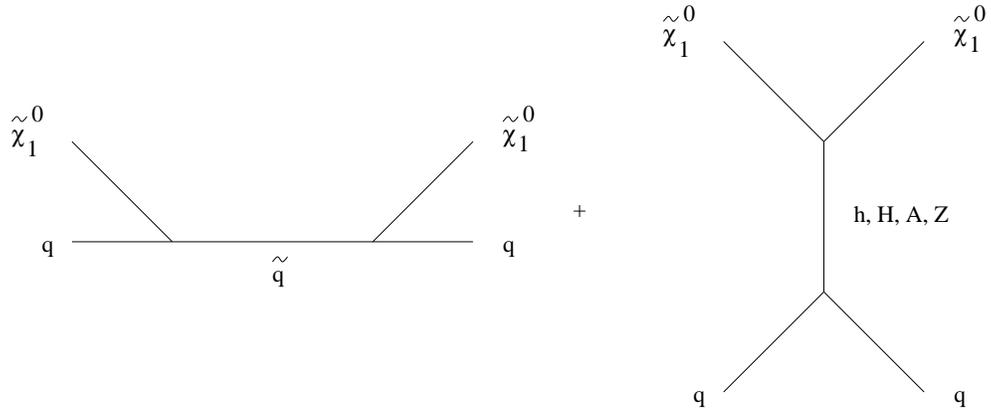 width 13 cm) }
\bigskip
\bigskip
\caption {\label{fig2} Scattering of $\tilde{\chi}_{1}^{0}$ by 
a quark of a nucleon through s channel 
$\tilde q$ and t channel $h$, $H$, $A$, $Z$ poles.}
\end{figure}

In general one sees that the bounds of Eq. (6) exert a strong inflence on the
allowed SUSY parameter space, and this in turn affects the size of the 
$\tilde\chi^0_1-q$ cross section. From Fig.~\ref{fig2} one has that 
(i)$\sigma_{\tilde\chi^0_1-p}$ becomes large for light (first generation) squarks
(i.e. small $m_0$) and light Higgs  bosons, which is just the region of rapid
early universe annihilation. Thus the lower bound on 
$\Omega_{\tilde\chi^0_1}h^2$ can produce an upper bound on
$\sigma_{\tilde\chi^0_1-p}$. (ii) We also note that $\sigma_{\tilde\chi^0_1-p}$
increases with tan$\beta$.

\section{mSUGRA Models} The mSUGRA model with radiative breaking of $SU(2)\times
U(1)$ depends upon four parameters and one sign. These may be taken as
following: m$_0$, the universal scalar mass at M$_G$; m$_{1/2}$, the universal
gaugino mass at M$_G$ (or alternately one may use $m_{\tilde\chi^0_1}$ or m$_{\tilde g}$
at the electroweak scale since these scale with m$_{1/2}$ i.e.
$m_{\tilde\chi^0_1}\simeq 0.4 m_{1/2}$ and $m_{\tilde g}\simeq 2.8 m_{1/2}$);
A$_0$, the universal soft breaking mass at M$_G$; tan$\beta=<H_2>/<H_1>$; and the
sign of $\mu$, the Higgs mixing parameter in the superpotential ($W_{\mu}=\mu
H_1 H_2$). (With our choice of sign in $W_{\mu}$, the $b\rightarrow s\gamma$
constraint eliminates most of the $\mu>0$ parameter space.)

\begin{figure}[htb]
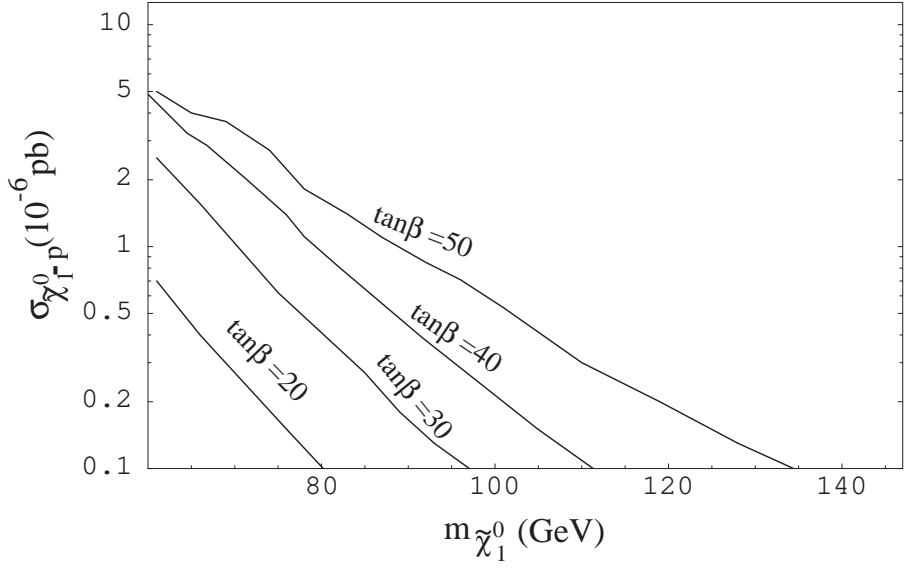

\centerline{ \DESepsf(aads20304050.epsf width 12 cm) }
\bigskip
\bigskip
\caption {\label{fig3} ( $\sigma_{\tilde{\chi}_{1}^{0}-p}$)$_{\rm
max}$
vs. $m_{\tilde{\chi}_{1}^{0}}$ 
for (from top to bottom) $\tan{\beta}=50$, 40, 30, and 20.}
\end{figure}

We proceed by varying the above parameters over the allowed space subject to all
the constraints discussed above. The maximum value of
$\sigma_{\tilde\chi^0_1-p}$ as a function of $m_{\tilde\chi^0_1}$ is given in
Fig.~\ref{fig3}. We see that current experiments sensitive to
$\sigma_{\tilde\chi^0_1-p}>1\times 10^{-6}$ pb are probing only large tan$\beta$
regime for this model, i.e. tan$\beta\stackrel{>}{\sim}25$.

\begin{figure}[htb]
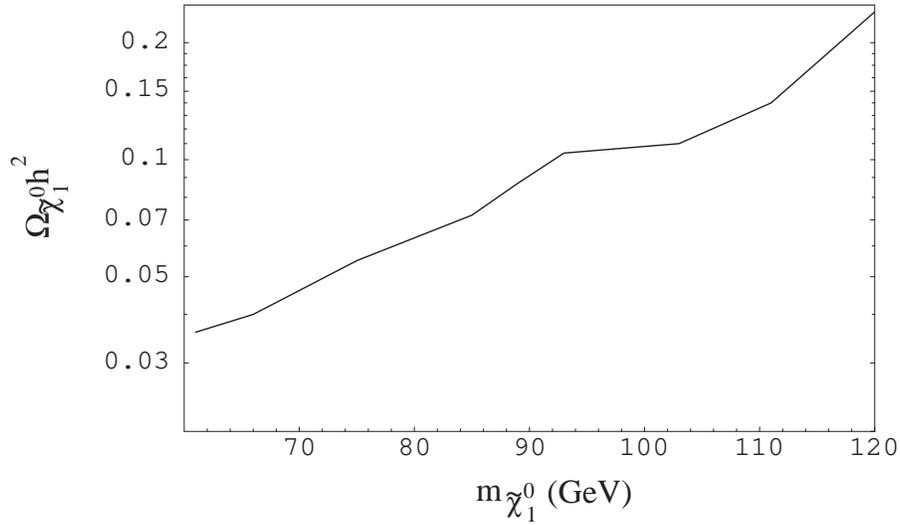

\centerline{ \DESepsf(aads30om.epsf width 12 cm) }
\bigskip
\bigskip
\caption {\label{fig4} Value of $\Omega_{\tilde{\chi}_{1}^{0}}h^{2}$
vs. $m_{\tilde{\chi}_{1}^{0}}$ for $\tan{\beta}=30$ when
$\sigma_{\tilde{\chi}_{1}^{0}-p}$ takes on its maximum value.}
\end{figure}

To obtain further insight as to the parameter space being probed by current
experiments, we show in Fig.~\ref{fig4} the value of $\Omega_{\tilde\chi^0_1}h^2$  when $\sigma_{\tilde\chi^0_1-p}$ takes on its
maximum value for the
characteristic case of tan$\beta$=30. We see that $\Omega_{\tilde\chi^0_1}h^2$ increases as
$m_{\tilde\chi^0_1}$ increases (as expected from the general discussion above)
from roughly the minimum to maximum value allowed by Eq.(6). Thus an accurate
determination of $\Omega_{m}h^2$ as might be expected from the MAP sattelite,
would significantly help narrow the allowed SUSY parameter space.

\begin{figure}[htb]
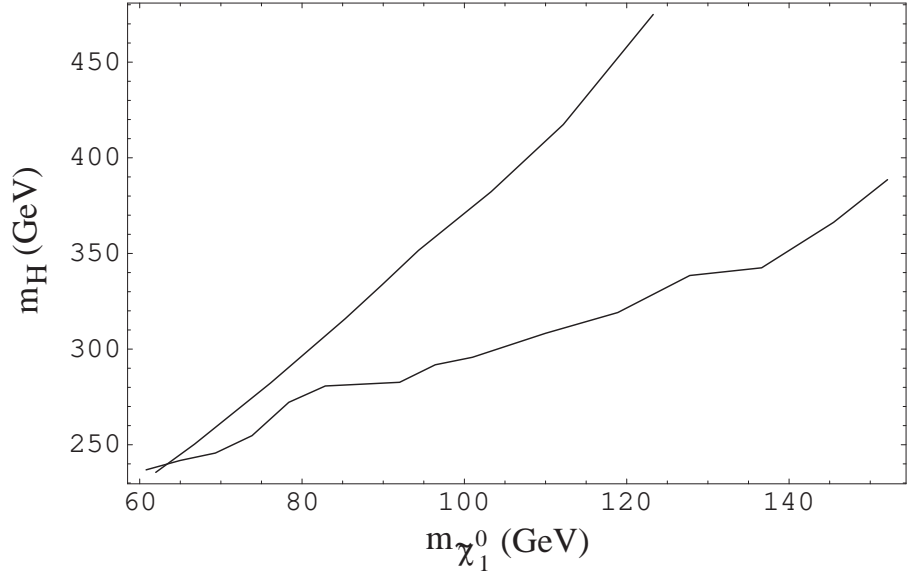

\centerline{ \DESepsf(aads3050Higgs.epsf width 12 cm) }
\bigskip
\bigskip
\caption {\label{fig5} $m_{H}$  when
$\sigma_{\tilde{\chi}_{1}^{0}-p}$ takes on its maximum value for
$\tan{\beta}=30$ (top curve) and  $\tan{\beta}=50$ (lower curve).}
\end{figure}

One notes in Fig. 3 that the larger tan$\beta$ cross section sustain with
increasing $m_{\tilde\chi^0_1}$ more than the
smaller tan$\beta$ do. This is due to the fact noted above that m$_H$ and m$_A$
become lighter as tan$\beta$ increases increasing $\sigma_{\tilde\chi^0_1-p}$.
This is shown in Fig.~\ref{fig5} for the two cases of tan$\beta$=30 and tan$\beta$=50. In
contrast, m$_h$ is relatively heavy, i.e. for tan$\beta$=30 one has that m$_h$
increases monotonically with $m_{\tilde\chi^0_1}$ from 112 GeV to 128 GeV. Since
m$_0$ is not large when $\sigma_{\tilde\chi^0_1-p}$ is at maximum, the squarks
lie below and close to the gluino where $m_{\tilde g}\simeq 7
m_{\tilde\chi^0_1}.$

\section{Nonuniversal Models}

The possibility of nonuniversal soft breaking significantly changes the region
of parameter space being accessed by current dark matter experiments. We
consider here the case where nonuniversal soft breaking masses are allowed both
in the Higgs and third generation sectors. A general parametrization at $M_G$
then 
\begin{eqnarray}
m_{H_{1}}^{\ 2}&=&m_{0}^{2}(1+\delta_{1}); 
\quad m_{H_{2}}^{\ 2}=m_{0}^{2}(1+ \delta_{2});\\\nonumber
m_{q_{L}}^{\ 2}&=&m_{0}^{2}(1+\delta_{3}); \quad m_{u_{R}}^{\ 2}=m_{0}^{2}(1+\delta_{4});
\quad m_{e_{R}}^{\ 2}=m_{0}^{2}(1+\delta_{5}); 	\\\nonumber
m_{b_{R}}^{\ 2}&=&m_{0}^{2}(1+\delta_{6}); \quad m_{l_{L}}^{\
2}=m_{0}^{2}(1+\delta_{7}).	\label{eq18}
\end{eqnarray}
Here $m_{0}$ is the universal mass of the first two generations,
 and $\delta_i$ are the deviations for the Higgs  and third generation. We
 assume here that $-1\leq \delta_i\leq 1$. (Note that for $SU(5)$ one would have
 $\delta_3=\delta_4=\delta_5$ and $\delta_6=\delta_7$.)
 
In order to understand the effects that occur in this more complicated
situation we first note thate $\tilde\chi^0_1$ is a mixture of gaugino and
higgsino  part : $\tilde\chi^0_1=\alpha\tilde W_3+\beta\tilde B+\gamma\tilde
H_1+\delta\tilde H_2$. Further, the spin independent cross section arises due
to the interference between the gaugino and higgsino parts of $\tilde\chi^0_1$,
causing 
$\sigma_{\tilde\chi^0_1-p}$ to increase with increasing interference. The SUSY
parameter that to a large extent controls the amount of interference is $\mu^2$,
interference increasing (and hence $\sigma_{\tilde\chi^0_1-p}$ increasing) as
$\mu^2$ decreases, and interference decreases as $\mu^2$ increases. The value of
$\mu^2$ at the electroweak scale in terms of the GUT scale parameter is
determined by the RGEs. In general, one must solve these numerically. However,
one can get a qualitative understanding of the effects of the $\delta_i$ from an analytic
solution which is valid for low and intermediate tan$\beta$\cite{nath1}:
\begin{eqnarray}
\mu^2&=&{t^2\over{t^2-1}}[({{1-3 D_0}\over 2}-{1\over
t^2})+({{1-D_0}\over2}(\delta_3+\delta_4)-{{1+D_0}\over2}\delta_2+{\delta_1\over
t^2})]m_0^2\\\nonumber &+& {\rm {universal \, parts +loop \, corrections}}.
\end{eqnarray}
Here $t\equiv tan\beta$, and $D_0\simeq 1-(m_t/200 sin\beta)^2$. In general
$D_0$ is small i.e $D_0\leq 0.23$. Eq. (9) shows  that it is necessary to
consider both the squark nonuniversalities ($\delta_3$ and $\delta_4$) as well as
the Higgs ($\delta_1$ and $\delta_2$) since they produce effects of comparable
size.

\begin{figure}[htb]
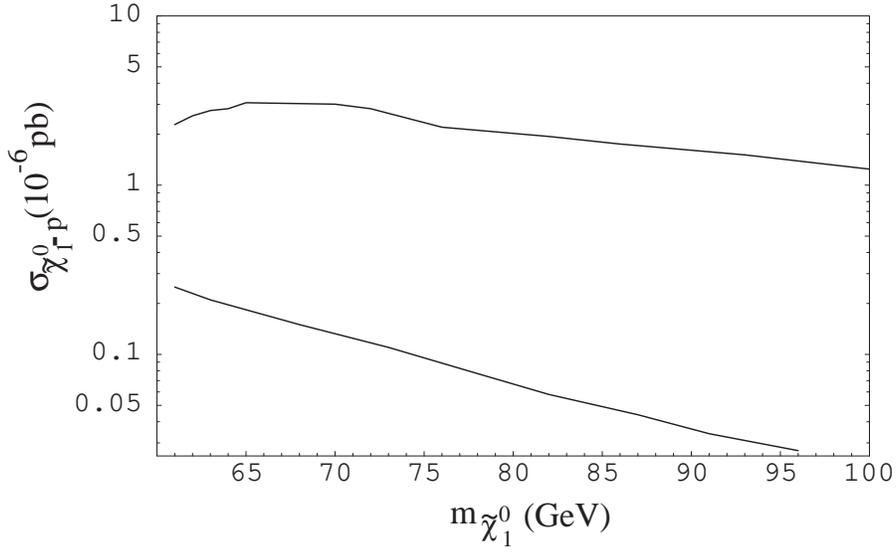

\centerline{ \DESepsf(aads7uni7nonuni.epsf width 12 cm) }
\bigskip
\bigskip
\caption {\label{fig6} Maximum 
 $\sigma_{\tilde{\chi}_{1}^{0}-p}$ 
 for $\tan{\beta}=7$ for mSUGRA (lower curve) and nonuniversal
 model (upper curve).}
\end{figure}

\begin{figure}[htb]
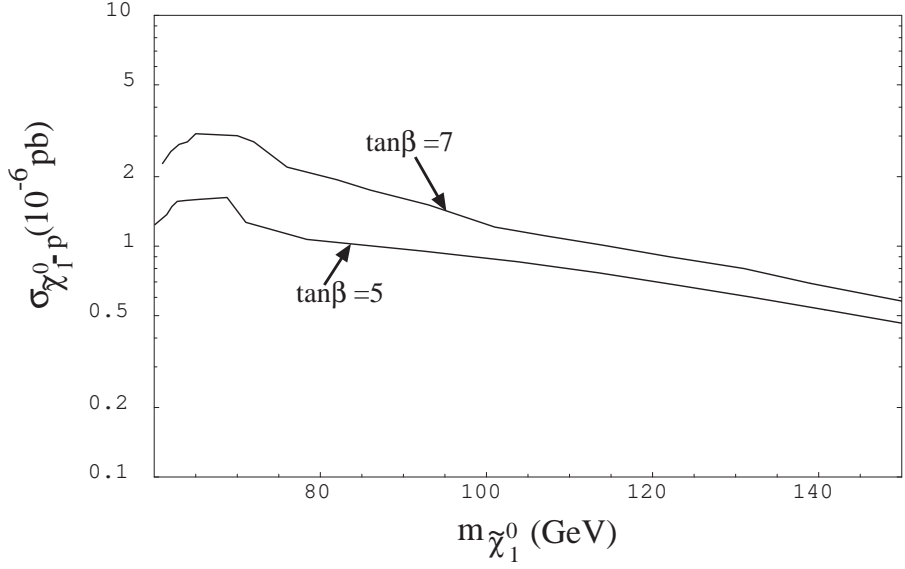

\centerline{ \DESepsf(aads57nonuni.epsf width 12 cm) }
\bigskip
\bigskip
\caption {\label{fig7} Maximum $\sigma_{\tilde{\chi}_{1}^{0}-p}$ 
for nonuniversal models for $\tan{\beta}=7$ 
(upper curve) and $\tan{\beta}=5$ (lower curve).}
\end{figure}

One sees from Eq. (9) that $\mu^2$ will be significantly reduced (and hence 
$\sigma_{\tilde\chi^0_1-p}$ increased) if one chooses $\delta_3,\delta_4,
\delta_1 <0$ and $\delta_2>$0. (The reverse will be the case for the opposite
choice of signs.) This effect can be seen in Fig.~\ref{fig6}, where the maximum value of 
$\sigma_{\chi^0_1-p}$ is plotted for tan$\beta=7$ for the nonuniversal model
(upper curve) and mSUGRA (lower curve). One sees that with the choice $\delta_3,\delta_4,
\delta_1 <0$ and $\delta_2>$0 one can increase the cross section by a factor of
10 to 100. As a consequence, there are regions of parameter space where the
nonuniversal models can be probed to much lower tan$\beta$. This is exhibited in
Fig.~\ref{fig7}. where the maximum cross section is given for tan$\beta$=7 and 
tan$\beta$=5. One sees that current experiments are probing tan$\beta$ as low as
tan$\beta\simeq$4. Since these experiments have also excluded the region 
$\sigma_{\tilde \chi^0_1-p}\stackrel{>}{\sim}10\times 10^{-6}$ pb, one finds that
part of the parameter space for tan$\stackrel{>}{\sim} 12$ is already
experimentally excluded for these models. Of course, the reverse choice of signs
for $\delta_i$ will lower the cross section for any value of tan$\beta$, leaving
such parts of the parameter space mostly as yet unexplored experimentally.

When  $\sigma_{\tilde \chi^0_1-p}$ takes on its maximum value for this model, the light Higgs becomes quite light lying just above the LEP 200 limit.
Thus for tan$\beta$=7 one finds 100 GeV$\leq m_h\leq$ 107 GeV which would make
the h relatively easy to find at the LHC and the Tevatron RUN II. In contrast
the squark can become quite heavy. This can be seen from  Eq. (9) where the
negative nonuniversal term dominates the $m_0^2$ contribution, and hence 
making $m_0^2$ large reduces $\mu^2$, making $\sigma_{\tilde \chi^0_1-p}$
larger. Thus the first two generation squarks for this situation have masses in
the range 600 GeV$\leq m_{\tilde q}\leq$ 1200 GeV, and lie above the gluino.
Large values of $m_{\tilde q}$ tend to suppress proton decay amplitudes of GUT
models and would help relieve the tension between a large $\sigma_{\tilde
\chi^0_1-p}$ and a small p decay rate.

\section{D-Brane Models}

Recent advances in string theory based on Dp-branes (manifolds of p+1
dimensions) has led to a revival of phenomenologically motivated string models.
One class of such models making use of type IIB orientifolds\cite{munoz} allows one
to put the Standard Model gauge group on 5-branes, manifolds of six dimensions, of
which four are usual Minkowski space and two are compactified on a torus.

An interesting model of this class puts $SU(3)_c\times U(1)_Y$ on one set of 5
branes, 5$_1$ and $SU(2)_L$ on a second intersecting set 5$_2$\cite{brhlik2}. Strings starting
on $5_2$ and ending on $5_1$ have massless modes carrying the joint quantum
numbers of the two branes (presumably the SM quark and lepton doublets and the
Higgs doublets), while strings begining and ending on 5$_1$ have modes carrying 
$SU(3)_C\times U(1)_Y$ quantum numbers (the right handed quark and
lepton states). This leads to a model with a unique set of nonuniversal gaugino
and squark and slepton masses at $M_G$ parametrized as follows:
\begin{eqnarray}
\tilde m_1&=&\tilde m_3=-A_0=\sqrt{3} cos\theta_b \Theta_1
e^{-i\alpha_1}m_{3/2}\\\nonumber
\tilde m_2&=&\sqrt{3} cos\theta_b (1-\Theta_1^2)^{1/2}m_{3/2}
\end{eqnarray}
and 
\begin{eqnarray}
m_{12}^2&=&(1-3/2 sin^2\theta_b)m_{3/2}^2\\\nonumber
m_{1}^2&=&(1-3 sin^2\theta_b)m_{3/2}^2
\end{eqnarray}
Here $\tilde m_i$, i=1,2,3 are three gaugino masses at $M_G$ , while $m_{12}^2$ are
$q_L$, $l_L$, $H_1$, $H_2$ sfermion and Higgs (mass)$^2$ and $m_1^2$ are 
$\tilde u_R$, $\tilde d_R$, $\tilde e_R$ sfermion (mass)$^2$. In addition, one has the B
soft breaking mass and the $\mu$ parameter at the GUT scale 
\begin{eqnarray}
B_0&=&|B_0|e^{i\theta_{0B}};\,\, \mu_0=|\mu_0|e^{i\theta_{0\mu}}
\end{eqnarray}
where $\alpha_1$, $\theta_{0B}$ and $\theta_{0\mu}$ are possible CP violating
phases. As can be seen from Eqs.(9,10), the parameters $\theta_b$ and $\Theta_1$
are restricted to $sin\theta_b\leq 1/\sqrt {3}$ ($\theta_b\leq 0.615$) and
$\Theta_1\leq 1$. 

Models of this type are of interest in that they are natural possibilities in
string theory, and yet it would be difficult to see how such a symmetry breaking
pattern could arise in conventional SUGRA GUT models. One aspect of the model is
that the cancelations between the gluino and neutralino CP violating phases in
the electron and neutron electric dipole moments (EDMs) implied by Eq. (10)
allows the CP violating phases  to be larger than usual at the elctroweak scale
and still satisfy the experimental EDM bounds \cite{de}. However, these
experimental bounds combined with the requirement of radiative breaking of
$SU(2)\times U(1)$ at the elctroweak scale, leads to serious fine tuning of
parameters at the GUT scale unless tan$\beta\stackrel{<}{\sim} 3-5$\cite{aad}. Since we will be
interested here in large tan$\beta$ (to investigate the maximum values of
$\sigma_{\chi^0_1-p}$) in the following we will first set all the CP violating
phases to zero. One then has a model with four parameters and one sign i.e.
$m_{3/2}$, $\theta_b$, $\Theta$ and tan$\beta$, and the sign of $\mu$. (As
before, $\mu$ must be negative for most of the parameter space to satisfy the
$b\rightarrow s\gamma$ constraint.) 

\begin{figure}[htb]
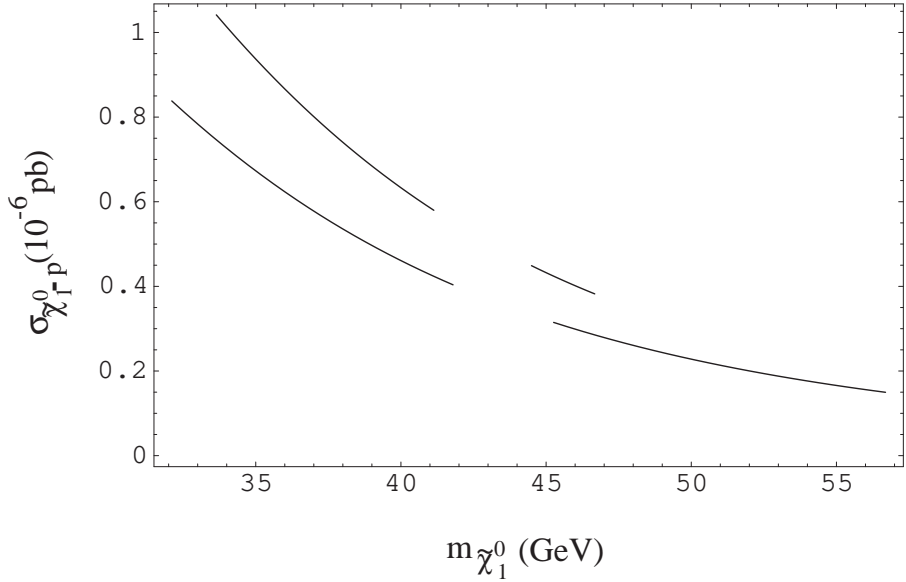

\centerline{ \DESepsf( darktalk20001.epsf width 12 cm) }
\caption {\label{fig8} $\sigma_{\tilde{\chi}_{1}^{0}-p}$ 
for the D-brane model for $m_{3/2}=175$ GeV, tan$\beta$=10 for
$\theta_b$=0.5 (upper curve), $\theta_b$=0.2 (lower curve). (The gaps are
excluded regions.)}
\end{figure}

\begin{figure}[htb]
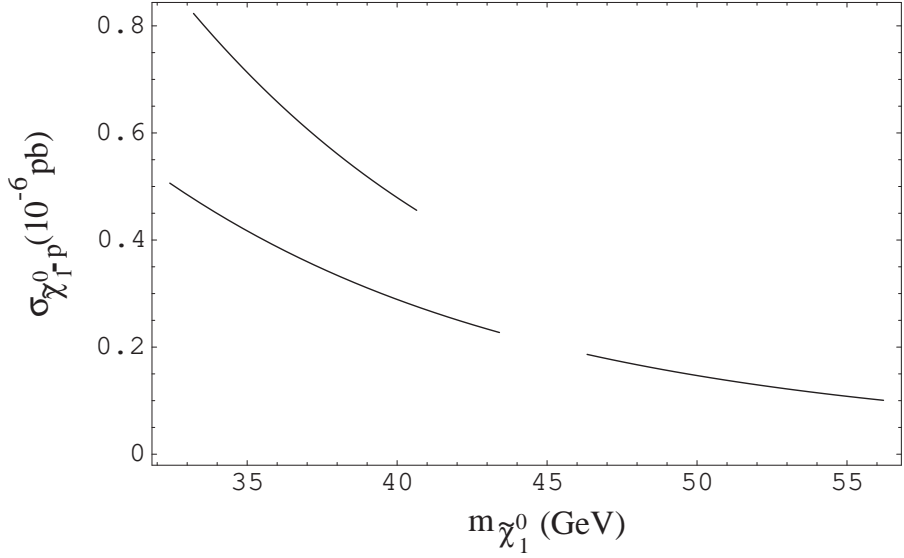

\centerline{ \DESepsf( darktalk20002.epsf width 12 cm) }
\caption {\label{fig9} $\sigma_{\tilde{\chi}_{1}^{0}-p}$ 
for the D-brane model for $m_{3/2}=200$ GeV, tan$\beta$=15 for
$\theta_b$=0.5 (upper curve), $\theta_b$=0.2 (lower curve).}
\end{figure}

Fig.~\ref{fig8} shows $\sigma_{\tilde\chi^0_1-p}$ as a function of $m_{\tilde \chi^0_1}$, 
which is parametrized by $\theta_b$, $\Theta_1$, $m_{3/2}$.
One has that $\sigma_{\tilde\chi^0_1-p}$ increases with increasing $\theta_b$ since
by Eq. (10) the squark masses of Fig.~\ref{fig2} decrease.
Fig.~\ref{fig9} shows $\sigma_{\tilde\chi^0_1-p}$ for tan$\beta$=15 and $m_{3/2}$=200 GeV. One
sees that one needs a large tan$\beta$, i.e. $tan\beta\stackrel{>}{\sim} 15$ to obtain a cross
section within current experimental sensitivities, i.e. $\sigma_{\tilde\chi^0_1-p}
\stackrel{>}{\sim}1\times 10^{-6}$ pb.
 
\begin{figure}[htb]
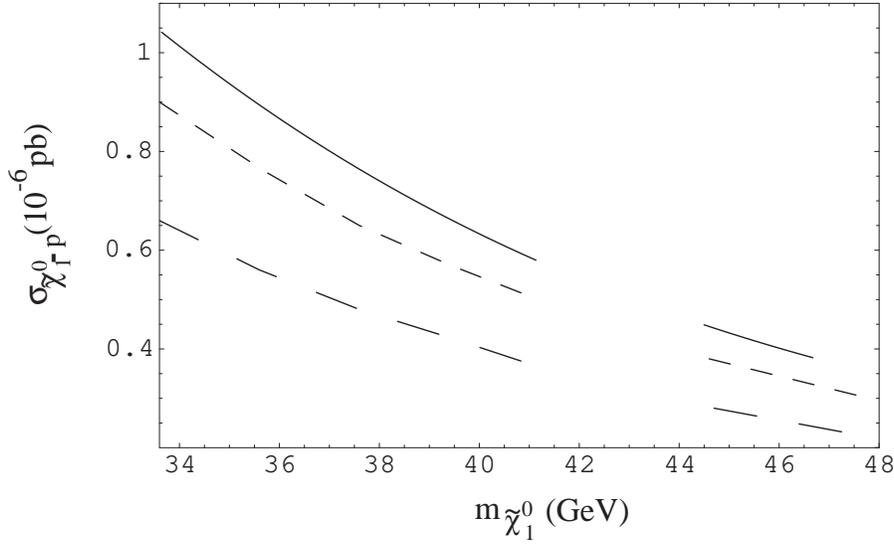

\centerline{ \DESepsf( darktalk20003.epsf width 12 cm) }
\bigskip
\bigskip
\caption {\label{fig10} $\sigma_{\tilde{\chi}_{1}^{0}-p}$
 for $\theta_b$=0.4, $m_{3/2}=200$ GeV, tan$\beta$=15 for
$\alpha_1=0$, $1.85\pi$, $1.75\pi$ in decreasing order.}
\end{figure}

\begin{figure}[htb]
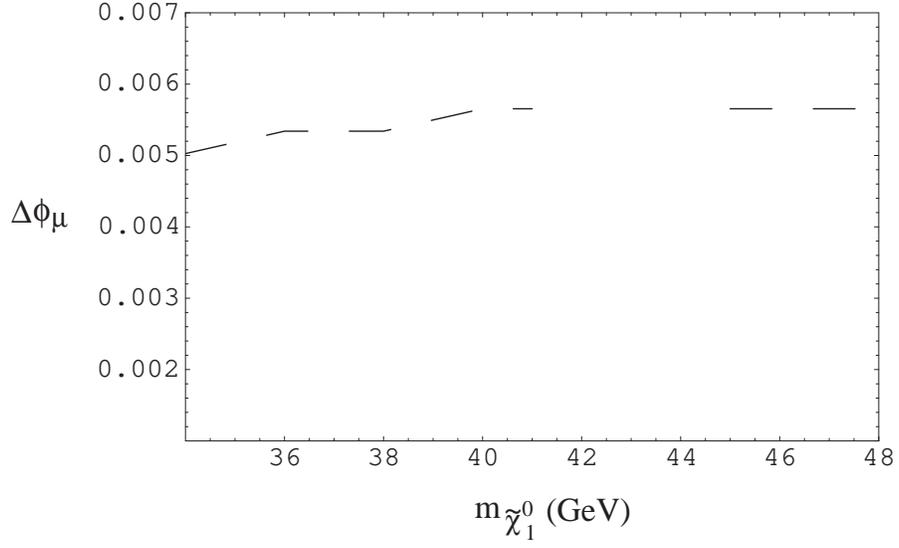

\centerline{ \DESepsf( darktalk20004.epsf width 12 cm) }
\bigskip
\bigskip
\caption {\label{fig11} The allowed range $\Delta\phi_{\mu}$ of the phase
$\phi_{\mu}$ for $\theta_b$=0.4, $m_{3/2}=200$ GeV, tan$\beta$=15 for
$\alpha_1=1.75\pi$.}
\end{figure}

 We next allow the CP violating phases to be non-zero to investigate their
 effect on $\sigma_{\tilde \chi^0_1-p}$ cross section. We impose here the experimental
 constraint on the electron EDM of $d_e< 4.3\times 10 ^{-27}$ ecm at 95$\%$ C.L.
 \cite{de}. (We do not impose the neutron EDM constraint here as there are a
 number of ambiguities in calculating $d_n$ \cite{aad}). Fig.~\ref{fig10} shows that 
 $\sigma_{\tilde \chi^0_1-p}$ decreases with increasing phase 2$\pi-\alpha_1$, the
 presence of the phase decreasing $\sigma_{\chi^0_1-p}$, by a factor of two more.
 As mentioned above, such large phases, while still satisfying the experimental
 bound on $d_e$, require a serious fine tuning of other parametrs. 
 Fig.~\ref{fig11} shows
 the allowed range $\Delta \phi_\mu$ of the phase $\phi_\mu$ to satisfy the
 electron EDM for tan$\beta$=15, $\alpha_1=1.75 \pi$. One sees that
 $\phi_{\mu}$ must be chosen very precisely to satisfy the EDM constraint on
 $d_e$. If one were also to impose the neutron EDM constraint as well, the fine
 tuning would be even more severe \cite{ads1}.
 \section{Conclusions}
 We have considered here the question of what part of the SUSY parameter space
 can be probed by detectors sensitive to the neutralino proton cross section in the
 range of Eq. (1). In order to examine this we have considered the maximum
 theoretical cross section as function of $m_{\tilde \chi^0_1}$. The answer depends on
 the particular SUSY model one is considering. Thus for mSUGRA models, one
 requires tan$\beta\stackrel{>}{\sim}$ 25 to achieve current detector sensitivities
 $\sigma_{\tilde \chi^0_1-p}\geq 1\times 10^{-6}$ pb. In this case, a large 
 $\sigma_{\tilde \chi^0_1-p}$ corresponds to relatively heavy Higgs, $m_h\simeq
 (110-130)$ GeV, a light neutralino of $m_{\tilde
 \chi^0_1}\stackrel{<}{\sim}120$ GeV (from the relic density constraint) and
 moderate squark masses (e.g. $m_{\tilde d}\simeq (400-700)$ GeV lying below but
 close to the gluino mass.
 
 Nonuniversal SUGRA models can increase or decrease $\sigma_{\tilde \chi^0_1-p}$
 by a factor of 10 to 100. In the former case, one can begin to probe the
 parameter space for tan$\beta\stackrel{>}{\sim}$ 4 for $\sigma_{\tilde
 \chi^0_1-p}\geq 1\times 10^{-6}$ pb. Here the Higgs is relatively light,
 $m_h\simeq (100-110)$ GeV, while the squarks are heavy (e.g. $m_{\tilde d}\simeq
 600-1200$ GeV) and lie well above the gluino. Current data has in fact begun to
 eliminate part of the parameter space for tan$\beta\stackrel{>}{\sim}$ 15.
 
 In the D-brane model considered, one requires tan$\beta\stackrel{>}{\sim}$ 15
 for $\sigma_{\tilde \chi^0_1-p}\geq 1\times 10^{-6}$ pb. CP violating phases
 that can appear in such models lower the $\tilde\chi^0_1-p$ cross section by a
 factor of two or more. However, there is a serious fine tuning problem for the
 $\mu$ phase at $M_G$ for such large values of $\tan\beta$.
 
 The fact that in the nonuniversal SUGRA models one can get a large
 $\sigma_{\tilde \chi^0_1-p}$ with a small tan$\beta$ and large squark mass
 tends to relieve some of the tension between dark matter cross sections and
 proton decay. Thus if $\tilde \chi^0_1$ dark matter were discovered with 
 $\sigma_{\tilde \chi^0_1-p}\stackrel{>}{\sim}1\times 10^{-6}$ pb, such models
 could still have a low proton decay rate since $\tau_p$ is suppressed for 
 small $\tan\beta$ and large $m_{\tilde q}$.
 
 In our analysis here, we have neglected the possibility of coannihilation
 effects since the rapid early universe annihilation produced by such effects
 generally require raising $m_0$ and $m_{1/2}$ to avoid violating the lower
 bound of Eq. (6), lowering the $\tilde \chi^0_1-p$ cross section . (We have
 been intersted here in the maximum values of $\sigma_{\tilde \chi^0_1-p}$.) We
 note, however, that aside from the coannihilation region arising from the
 $\tilde \tau-\tilde \chi^0_1$ degeneracy for mSUGRA at low and intermediate
 tan$\beta$ discussed in \cite{ellis1}, there are three other coannihilation regions
 for the models considered here. In mSUGRA with large tan$\beta$, the chargino
 $\tilde \chi^{\pm}_1$ can become degenerate with the $\tilde \chi^0_1$ leading
 to a new coannihilation domain. In the nonuniversal SUGRA model, one can see
 from Eq. (8) that the choice $\delta_5<0$ can make the $\tilde\tau_R$ light and
 hence become degenerate with the $\tilde \chi^0_1$ even for small tan$\beta$.
 For the D-brane models one sees from Eq. (9) that as $\Theta_1$ gets large,
 $\tilde m_2$ and $\tilde m_1$ become  equal allowing the $\tilde
 \chi^{\pm}_1$ to become degenerate with $\tilde \chi^0_1$ (roughly
 for $\Theta_1\stackrel{>}{\sim}0.8$). Unlike the case considered in \cite{ellis1},
 each of these new domains can occur for light neutralinos where $\sigma_{\tilde
 \chi^0_1-p}$ may be large. Coannihilation effect will be discussed elsewhere
 \cite{ads2}.
 
\section{Acknowledgments}

This work was supported in part by National Science Foundation Grant No.
PHY-9722090.



\end{document}